\documentclass{emulateapj}
\usepackage{apjfonts}
\usepackage{graphicx}

\usepackage{amsmath}

\shorttitle{TWO-STEP ACCELERATION MODEL OF COSMIC RAYS AT MIDDLE-AGED SNR}
\shortauthors{T. INOUE ET AL. }

\begin{document}

\title{
TWO-STEP ACCELERATION MODEL OF COSMIC RAYS AT MIDDLE-AGED SUPERNOVA REMNANTS: UNIVERSALITY IN SECONDARY SHOCKS
}
\author{Tsuyoshi Inoue\altaffilmark{1,2}, Ryo Yamazaki\altaffilmark{2}, and Shu-ichiro Inutsuka\altaffilmark{3}}
\altaffiltext{1}{Division of Theoretical Astronomy, National Astronomical Observatory of Japan, Osawa, Mitaka 181-8588 Japan; inouety@th.nao.ac.jp}
\altaffiltext{2}{Department of Physics and Mathematics, Aoyama Gakuin University, Fuchinobe, Sagamihara 229-8558, Japan}
\altaffiltext{3}{Department of Physics, Nagoya University, Nagoya 464-8602, Japan}

\begin{abstract}
Recent gamma-ray observations of middle-aged supernova remnants revealed a mysterious broken power-law spectrum.
Using three-dimensional magnetohydrodynamics simulations, we show that the interaction between a supernova blast wave and interstellar clouds formed by thermal instability generates multiple reflected shocks.
The typical Mach numbers of the reflected shocks are shown to be $M\simeq$ 2 depending on the density contrast between the diffuse intercloud gas and clouds.
These secondary shocks can further energize cosmic-ray particles originally accelerated at the blast-wave shock.
This ``two-step" acceleration scenario reproduces the observed gamma-ray spectrum and predicts the high-energy spectral index ranging approximately from 3 to 4.
\end{abstract}


\section{Introduction}
Supernova remnants (SNRs) are believed to be the sites of Galactic cosmic-ray acceleration through a diffusive shock acceleration mechanism (DSA; see, e.g., Blandford \& Eichler 1987) and multi-wavelength nonthermal emissions from SNRs originating in accelerated particles have been detected (Koyama et al. 1995, Aharonian et al. 2008a).
Recent observations using Fermi Gamma-ray Space Telescope revealed the gamma-ray spectra from middle-aged SNRs; W51C, W28, and W44 (Abdo et al. 2009b; 2010c; 2010d).
The observed gamma-ray spectra is likely to be explained by accelerated nuclei whose energy distributions are universally described by a broken power-law with an break energy of approximately $10$ GeV.
These SNRs are also known to be interacting with molecular clouds.
In a neutral medium such as in the vicinity of molecular clouds, the damping of magnetohydrodynamic (MHD) waves by ion-neutral collisions leads to escape of particles from the shock wave that interrupts acceleration at some break energy (Abdo et al. 2009b).
Theoretical studies of the particle acceleration that takes into account the effect of the wave dumping well explain the observed break energy of approximately $10$ GeV for middle-aged SNRs (Ptuskin \& Zirakashvili 2003).
In order to understand the observed broken power-law spectrum, however, we should still find some mechanisms that accelerate particles beyond the break energy whose spectrum continues, at least, to the TeV energy level as a power-law.
The power-law index $s$ of the accelerated particles above the break point is roughly $s\sim 3$-$4$, which is much steeper than the conventional DSA scenario that predicts $s \simeq 2$.
So far several scenarios have been proposed to explain the observed spectra (Malkov et al. 2010; Ohira et al. 2010b; Uchiyama et al. 2010; Li \& Chen 2010).
In this Letter, we propose another one.

\section{Cloudy Interstellar Medium}
To study the physics of the shock-cloud interaction in SNRs, we examine the propagation of a shock wave through a medium generated as a natural consequence of the thermal instability.
The thermal instability is a mechanism for condensation of interstellar gas driven by runaway radiative cooling (Field 1965).
Theoretically, the thermal instability is shown to be effective during the cloud formation process (Koyama \& Inutsuka 2002; Hennebelle et al. 2008; Inoue \& Inutsuka 2008; 2009), and it always dominates the dynamics of the cloud formation (Heitsch et al. 2008).
In this work, the three-dimensional, ideal magnetohydrodynamics equations including the effects of radiative cooling, heating, and thermal conduction are solved using the Godunov-CMOC-CT scheme (Inoue et al. 2009; Sano et al. 1999; Clarke 1996).
We perform simulations in (2 pc)$^3$ numerical domain with a grid spacing $\Delta x=2/1024$ pc $\simeq 2\times 10^{-3}$ pc, imposing a periodic boundary condition.
With this resolution, we can safely express the smallest structure formed by the thermal instability (Koyama \& Inutsuka 2004; Inoue et al. 2006).
Panel (a) of Fig.~\ref{f1} shows the result of the cloud formation by the thermal instability (detailed dynamics of the cloud formation can be found in Inoue \& Inutsuka 2008; 2009; Inoue et al. 2009).
The regions in blue represent the condensed clouds ($n_{\rm c}\simeq 40$ cm$^{-3}$, $T_{\rm c}\simeq 90$ K), while the diffuse intercloud gas is depicted in yellow ($n_{\rm i}\simeq 0.7$ cm$^{-3}$, $T_{\rm i}\simeq 5000$ K).
We have initially imposed the $y$-directional uniform magnetic field with strength $B=5$ $\mu$G, which is typical in the interstellar medium (Bech 2000).
Since the most unstable scale of the thermal instability is $\sim 1$ pc and the clouds are formed by condensation along the magnetic field line, the resulting clouds have sheet-like morphology whose scale is $\sim 1$ pc and thickness is $\lesssim 0.1$ pc.
Observationally, the structures formed by the thermal instability are expected in the envelope of molecular clouds (Sakamoto \& Sunada 2003), where the interaction with a supernova shock wave takes place.

\section{Formation of SNR}
\begin{figure}[h]
\epsscale{1.7}
\plotone{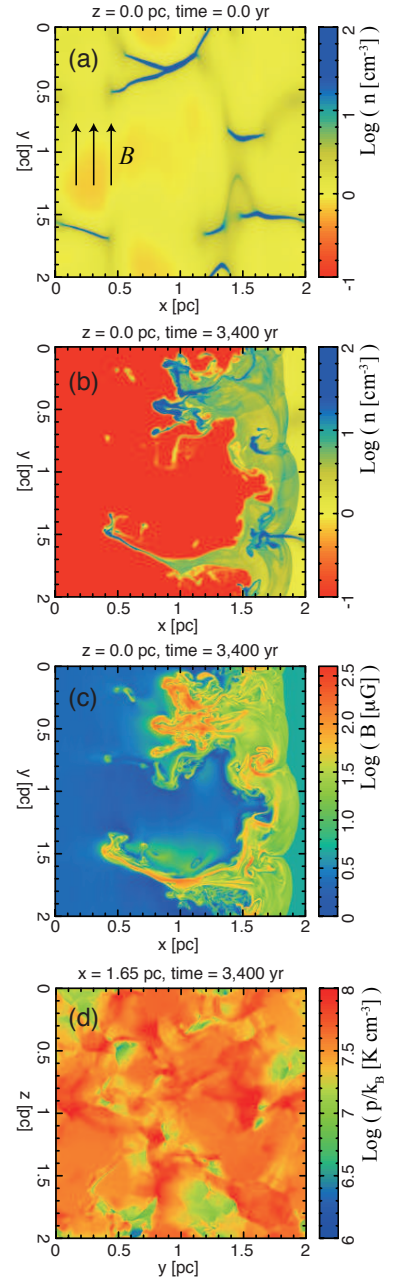}
\caption{
Panel (a) shows a two-dimensional slice of density distribution of the cloudy medium as a consequence of thermal instability.
Arrows in the panel indicate the orientation of the initial magnetic field.
Panel (b) is a number density slice after 3,400 years shock propagation.
Panel (c) is a slice of magnetic field strength at $t=3,400$ yr.
Panel (d) is a pressure slice at $t=3,400$ yr.
In the panels (a), (b), and (c), the $x$-$y$ plane at $z=0$ pc are shown.
In the panel (d), to display clearly the existence of the secondary shocks, the $y$-$z$ plane at $x=1.65$ pc is shown.
}
\label{f1}
\end{figure}

By setting static, high-pressure plasma with $P_{\rm h}/k_{\rm B}=5.0\times10^7$ K cm$^{-3}$ and $n_{\rm h}=0.1$ cm$^{-3}$ at the $x=0$ pc boundary, we inject a strong shock wave that propagates into the cloudy medium.
The resulting average value of the shock velocity is $536$ km s$^{-1}$, which corresponds to that of the SNR with an age of $\sim 10^4$ yr.
The density structure after the propagation of 3400 years is shown in the panel (b) of Fig.~\ref{f1}.
In principle, a shock wave hitting a dense cloud should be stalled, because the momentum conservation results in a smaller velocity for medium with increased inertia. 
However, the shock wave maintains high velocity in the diffuse intercloud medium that fills most of the volume.
Thus, we can expect the conventional picture of DSA in most of the volume, even if the SNR interacts with dense clouds.

Due to the velocity difference between the clouds and intercloud region, the shock is highly deformed.
It is known that, even if the pre-shock medium is uniform, the curved shock wave leaves vorticity behind the shock wave (Crocco's theorem), driving turbulence in the SNR.
As shown in the panel (c) of Fig.~\ref{f1}, in such a turbulent downstream region, the magnetic field is amplified by the stretching of the field lines (Giacalone \& Jokipii 2007; Inoue et al. 2009; see also Jones \& Kang 1993 who predicted the magnetic field amplification at the shear layer around the cloud caused by the shock-cloud interaction).
In the present case, the average magnetic field strength in the SNR saturates approximately at $\sim 40 \mu$G, which is twice the maximum field strength achieved by the shock compression alone.
The maximum magnetic field strength saturates at $\sim 400 \mu$G in the regions where local plasma $\beta$ is on the order of unity.

\section{Secondary Shocks}
When a shock wave hits a dense object, a reflected shock wave forms and propagates back into the post-shock medium, in addition to the transmitted shock wave.
In the present case, the pre-existing dense clouds in the pre-shock medium generate a number of reflected shock waves.
We call these reflected shock waves ``secondary shocks'' to contrast them with the original ``primary shock'' that propagates into the unshocked medium.
Panel (d) of Fig.~\ref{f1} shows the pressure distribution in the cross section of the SNR, indicating the existence of such secondary shock waves as discontinuous pressure jumps.
The strength of the secondary shocks can be estimated by analyzing the shock tube problem that describes the interaction between a dense gas and a pressure driven shock wave (Miesch \& Zweibel 1994).
The Mach number of the reflected shock wave $M_{\rm r}$ measured in the rest-frame of the shocked diffuse gas is obtained by solving a polynomial equation that is characterized by two parameters, namely the pressure ratio $\delta=P_{\rm c}/P_{\rm s}$ of the cloud and shocked gas and the density ratio $\epsilon=\rho_{\rm s}/\rho_{\rm c}$ of the shocked diffuse gas and the cloud (Miesch \& Zweibel 1994).
Fortunately, the parameter $\delta$ is very small ($\sim10^{-4}$) in the present simulation.
Expanding the polynomial with respect to $\delta$, we find
\begin{equation}
\left( M_{\rm r}^2-\frac{4}{\sqrt{5}}\,M_{\rm r}-1 \right)^{\!2}\! -\epsilon\,M_{\rm r}\!\left( M_{\rm r}^2-\frac{1}{5}\right)\!+\!O(\delta)=0,\label{Mr}
\end{equation}
where we have assumed the adiabatic index $\gamma=5/3$.
Substituting the typical density ratio $\epsilon =10^{-1}$ achieved in the simulation, we obtain $M_{\rm r}=1.80$.
In the limit of $\epsilon \rightarrow 0$ (i.e., the very dense cloud case), we obtain $M_{\rm r}\rightarrow\sqrt{5}$.
Because the primary shock is driven by isotropic thermal pressure and the transmitted shock wave is stalled heavily due to the large cloud density, the primary shock tends to hit the cloud perpendicular to its surface.
Thus, the above formula obtained by assuming one-dimensional geometry always enables us to evaluate $M_{\rm r}$, even though clouds have complex structure.

\begin{figure}[t]
\epsscale{1.}
\plotone{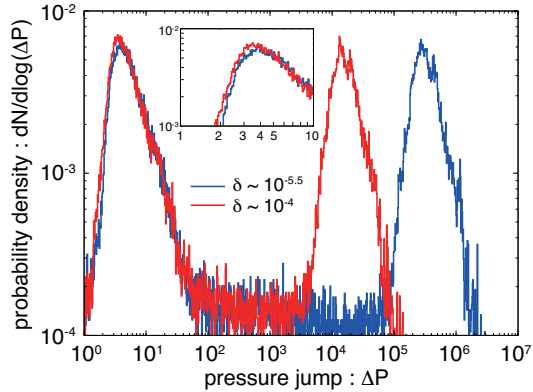}
\caption{
Probability distribution function (PDF) of pressure jump $\Delta P$.
Red and blue lines are the results of the simulations with the pressure ratio $\delta\sim10^{-4}$ and $10^{-5.5}$, respectively.
Inner panel is the close up of $\Delta P \in [1,10]$.
Integrated area of the PDF in a specific range of the pressure jump is proportional to the total surface area of the shock waves whose strengths are in the given range.
Thus, it is clear that the total surface areas of the secondary shocks is comparable to that of the primary shock.
}
\label{f2}
\end{figure}

To show the strengths of the shock waves in the simulation, we plot the probability distribution function (PDF) of the pressure jumps in Fig.~\ref{f2} as a red line.
Our algorithm to detect the pressure jump is as follows: 
We define the sphere whose radius is 20 times numerical grid size and whose center is at each cell in the numerical domain.
In each sphere, we calculate the minimum pressure $P_{\rm min}$, the maximum pressure $P_{\rm max}$, and the maximum pressure gradient, except for the injected hot plasma with $n<0.5$ cm$^{-3}$. 
If the maximum pressure gradient is larger than the critical pressure gradient $P_{\rm max}/(5\,\Delta x)$, i.e., the pressure fluctuation in the sphere is caused within the narrow region, we regard a shock wave as being in the sphere whose pressure jump is $\Delta P=P_{\rm max}/P_{\rm min}$.
Here the factor 5 in the expression of the critical pressure gradient is chosen from the fact that the high-resolution shock-capturing scheme used in this study requires 5 grid points at most to express a shock wave.
The bimodal distribution appearing in the PDF shows the existence of the primary and the secondary shocks.
According to the Rankine-Hugoniot relation, the Mach number of a shock for the gas of $\gamma=5/3$ is related to the pressure jump as (Landau \& Lifshitz 1959):
\begin{equation}\label{RH}
M(\Delta P)=\sqrt{ \frac{1}{5}+\frac{4}{5}\,\Delta P}.
\end{equation}
Substituting the peak of the PDF $\Delta P=3.85$ that corresponds to the secondary shocks, we obtain $M=1.81$, which agrees well with the above analytic evaluation.
To reinforce our analysis, we have performed the additional simulation with larger primary shock velocity with $v_{\rm sh}=2403$ km s$^{-1}$, i.e., the simulation with smaller $\delta$ ($\sim 10^{-5.5}$).
Since $M_{\rm r}$ is almost independent of the parameter $\delta$, we should obtain almost the same PDF of the secondary shocks.
The blue line in Fig.~\ref{f2} shows the result of the additional simulation verifying our analytic estimate for the strengths of the secondary shocks.

\section{Origin of Broken Power-Law Spectrum of Cosmic Rays}
As mentioned earlier, recent gamma-ray observations of middle-aged SNRs have revealed a steep particle spectrum with power-law index $s\simeq3$-$4$, while $s\simeq 2$ below $\sim10$ GeV.
In the following, we consider DSA in the environment obtained by our simulation to explain the observed spectrum.
Our idea is similar to that given in Jones \& Kang (1993) who pointed out that the reflected shock wave caused by the shock-cloud interaction can affect the cosmic-ray acceleration in the SNR.

The spectral index of the cosmic rays accelerated by the DSA mechanism at a shock wave with the Mach number $M$ is given by (Blandford \& Eichler 1987):
\begin{equation}\label{SDSA}
s=\frac{2\,(M^2+1)}{M^2-1}.
\end{equation}
Thus, the secondary shocks with $M=1.81$ achieved in the simulation can accelerate particles with the spectral index $s= 3.75$, while the primary shock with $M\gg10$ gives $s\simeq2.0$.
In the galactic interstellar medium, the densities of the intercloud gas and HI clouds are determined by the balance of radiative cooling and heating.
From the detailed evaluation of the heating/cooling rates, it is known that the density of intercloud gas is $n_{\rm i}\la 1$ cm$^{-3}$ and that of HI clouds is $n_{\rm c}\ga 10$ cm$^{-3}$ between which gas is thermally unstable (see, e.g., Wolfire et al. 1995).
Thus, the maximum value of the parameter $\epsilon$ can be evaluated as $\epsilon_{\rm max}=4\,n_{\rm i,max}/n_{\rm c,min}\simeq 0.4$ that leads to the upper bound of the spectral index $s\sim 4$ from eqs. (\ref{Mr}), (\ref{RH}) and (\ref{SDSA}).
If the preshock clouds are molecular and much denser than the HI cloud, the parameter $\epsilon$ can be substantially smaller than 0.1 realized in the present simulation.
From the typical Mach number of the secondary shocks in the dense cloud limit ($M=\sqrt{5}$), the lower bound of the index would be $s\simeq 3$.
Thus, the spectral index of the particles accelerated at the secondary shock would be limited in the range $3\la s \la 4$, which agrees well with the recent observations of middle-aged SNRs (Abdo et al. 2009b; 2010c; 2010d).
Note that recent numerical simulations have shown that the two-phase structure of clouds and diffuse intercloud gas is also generated even inside molecular clouds as a result of the thermal instability (see, e.g., Hennebelle et al. 2008).
This suggests that our scenario would be unchanged even if supernova explosion happens inside molecular clouds.

In contrast to the primary shock, the secondary shocks are located in the fully-ionized, postshock medium of the primary shock.
Thus, the acceleration process at the secondary shocks is free from the wave damping process due to the ion-neutral collisions, so that the acceleration beyond the break energy is possible.
The maximum energy of accelerated protons achieved at the secondary shock can be estimated using the DSA theory (Malkov \& Drury 2001):
\begin{eqnarray}\label{Emax}
\!\!\!\!E_{\rm max}&=&\frac{3}{10}\,\eta\,\frac{v_{\rm sh}^{2}\,e\,B\,t_{\rm age}}{c}\nonumber\\
&\simeq&10^{14}\!\left( \frac{\eta}{1} \right)\!\!\left( \frac{v_{\rm sh}}{500\,\mbox{km s}^{-1}}\right)^{\!\!2}\!\!\left( \frac{B}{40\,\mu\mbox{G}}\right)\!\!\!\left( \frac{t_{\rm age}}{10^4\,\mbox{yr}}\right)\!\mbox{eV},
\end{eqnarray}
where we have scaled using the average magnetic field strength in the simulated SNR ($\langle |B|\rangle\sim 40\,\mu$G), and the degree of the magnetic field fluctuations $\eta=\delta B^2/B^2$ has evaluated using the power spectrum of the magnetic field for the particles with energy near $E_{\rm max}$.
For the shock velocity, we have substituted the primary shock velocity, since the sound speed in SNR, which is approximately the velocity of the secondary shocks, is always comparable to the primary shock speed (more precisely, $v_{\rm sh, pri}=3\,c_{\rm s}/\sqrt{5}$ for the strong shock with adiabatic index $\gamma=5/3$).
As for the acceleration timescale, we have substituted the age of SNR.
The minimum lifetime of the secondary shocks is given by the radial sound crossing time of the SNR shell, which is typically one tenth of the age of SNR.
Thus the maximum attainable energy estimated in eq. (\ref{Emax}) can be smaller by an order of magnitude, although the secondary shocks do not always propagate radially.
Note that the spatial extension of each secondary shock is roughly given by the length of the clouds, which is essentially determined by the most unstable scale of the thermal instability $\sim 1$ pc.
Thus, the scale of each secondary shock is much larger than the gyro-radius of the maximum energy proton $l_{\rm g}=0.017\,(E/10^{14}\,{\rm eV})\,(B/40 \mu{\rm G})^{-1}$ pc, indicating that it does not restrict the maximum attainable energy.
Therefore, in principle, the secondary shocks can accelerate particles as large as 100 TeV in the middle-aged SNRs.
Even if $E_{\rm max}$ is an order of magnitude smaller owing to the lifetime of the secondary shocks, the maximum energy of 10 TeV is still large enough to account for the TeV gamma-ray emission from middle-aged SNRs (Abdo et al. 2009a; 2009b; 2010c; 2010d; Aharonian et al. 2002; 2008b; Buckley et al. 1998).

Since the Mach numbers of the secondary shocks are small, the injection rate of particles into the acceleration process in the secondary shocks would be much smaller than that in the primary shock.
However, suprathermal particles generated by the primary shock can be advected downstream and re-accelerated to higher energies at the secondary shocks.
As shown in Fig.~\ref{f2}, the total surface area of the secondary shocks is comparable to that of the primary shock, suggesting that most advected particles meet the secondary shocks.
If the volume filling factor of the clouds is much larger than that realized in our simulations, the total surface area of the secondary shocks would be much larger.
In that case the suprathermal particles can be successively accelerated by multiple secondary shocks.
Given the particle momentum spectrum accelerated at the primary shock as $N_{0}(p)\propto p^{-2}\,\exp(-p/p_{\rm br})$, where $p_{\rm br} \simeq 10$ GeV is the break momentum due to the wave damping, the spectrum re-accelerated $n$ times by the secondary shocks becomes (Melrose \& Pope 1993):
\begin{eqnarray}
\!\!N_{n}(p)&=&(s-1)\,(p/d)^{-s}\int_{0}^{p/d} q^{s-1}\,N_{n-1}(q/d)\,dq,\\
&=&p^{-2}\!_{n}G_{n}\!\!\left[a,...,a;\! b,...,b;\! \frac{-p}{p_{\rm br}\,d}\right]\!\!\left(\frac{b}{a}\right)^{\!\!n}\!\! d^{(n-2)s+2},\label{bpl}\\
&\propto& \bigg\{
\begin{array}{c}
p^{-2}\ (\mbox{for }p\ll p_{\rm br}), \\
p^{-s}\ (\mbox{for }p\gg p_{\rm br}), \\
\end{array}
\end{eqnarray}
where
\begin{equation}
_{m}G_{l}\left[g_1,...,g_m; h_1,...,h_l; z\right]=\sum^{\infty}_{k=0} \frac{(g_1)_{k}...(g_m)_{k}}{(h_1)_{k}...(h_l)_{k}}\frac{z^k}{k!},
\end{equation}
 is the generalized hypergeometric function written using the Pochhammer symbol [$(g)_k=g(g+1)...(g+k-1),\,(g)_0=1$], $a=s-2$, $b=s-1$, $d=\{(s-1)/(s+2)\}^{1/3}$.
 Here we have assumed that the injection at the secondary shocks is inefficient owing to their small Mach number.
Note that eq. (\ref{bpl}) is valid for $p_{\rm inj}<p< E_{\rm max}/c$, where $p_{\rm inj}$ is the injection momentum above which particles can cross the shock.
Moreover, we have assumed all the secondary shocks have the same Mach number and thus the same acceleration slope $s$ given by eq. (\ref{SDSA}).
Thus, we can expect the unique broken power-law spectrum irrespective of the experienced number of the re-acceleration.

\section{Predictions and Discussions}
The above results allow us to make the following predictions.
(a) The spectral index $s$ above the break energy cannot be substantially smaller than 3, since the typical Mach number of the reflected shocks in the dense cloud limit is $M_{\rm r}=\sqrt{5}$.
Note that this is not true for young SNRs, because thermal pressure in the Sedov-phase shell decreases toward its center, which makes eq. (\ref{Mr}) inapplicable.
In that case, the Mach number of the secondary shocks would increase as they propagate toward the center of the SNR, leading to a shallower slope.
The broken power-law spectra observed in a young SNR: Cassiopeia A (Abdo et al. 2010a) and possibly young SNR: IC443 (Abdo et al. 2010b) may be the case.
(b) Massive stars, progenitors of supernovae, are born in giant molecular clouds.
Thus, most of middle-aged SNRs in the galactic mid-plane will interact with molecular clouds, and will have a broken power-law cosmic-ray spectrum whose index above the break energy is bounded by $3 \la s \la 4$.
(c) If the spectral index above the break energy at a middle-aged SNR is substantially smaller than 3, the softening of the spectrum would be attributed to particle escape.
In that case the spectral index substantially depends on the evolution of maximum attainable energy $E_{\rm max}(t)$ and injection rate $\eta(t)$ of DSA (Ohira et al. 2010a; 2010b).
We can safely measure $E_{\rm max}(t)$ and $\eta(t)$ only in such SNRs, since the effect of the secondary shocks on the spectral slope would not be dominant there.

\acknowledgments We would like to thank Prof. T. Terasawa and Dr. Y. Ohira for fruitful discussions.
We are grateful for the anonymous referee whose helpful advices substantially improve the quality of this paper.
Numerical computations were carried out on XT4 at the Center for Computational Astrophysics (CfCA) of National Astronomical Observatory of Japan.
This work is supported by Grant-in-aids from the Ministry of Education, Culture, Sports, Science, and Technology (MEXT) of Japan, No.21740146 and No.22$\cdot$3369 (T. I.), No. 19047004 and No. 21740184 (R. Y.), and No. 16077202 and No. 18540238 (S. I.).

{}

\end{document}